\documentclass[letter]{aa} 

\usepackage{graphicx}
\usepackage{txfonts}
\usepackage[hidelinks]{hyperref}
\usepackage{amsmath}
\usepackage{amssymb}	
\usepackage[T1]{fontenc}
\usepackage{ae,aecompl}
\usepackage{multirow}
\usepackage{tabularx}
\usepackage{natbib}
\bibpunct{(}{)}{;}{a}{}{,} 

\usepackage{siunitx}

\begin{document}

   \title{Updated orbital ephemeris of the ADC source X 1822-371: a stable orbital expansion over 40 years}


   \author{S. M. Mazzola\inst{1}, R. Iaria\inst{1}, T. Di Salvo\inst{1}, A.F. Gambino\inst{2}, A. Marino\inst{1,3,4}, L. Burderi\inst{2}, A. Sanna\inst{2} \\ A. Riggio\inst{2} and M. Tailo\inst{5}}

   \institute{Dipartimento di Fisica e Chimica - Emilio Segrè,
 Universit\`a di Palermo, via Archirafi 36 - 90123 Palermo, Italy
         \and
         Dipartimento di Fisica, Universit\`a degli Studi di Cagliari, SP
Monserrato-Sestu, KM 0.7, Monserrato, 09042 Italy
   \and
  Istituto Nazionale di Astrofisica, IASF Palermo, Via U. La Malfa 153, I-90146 Palermo, Italy
  \and
  IRAP, Universitè de Toulouse, CNRS, UPS, CNES, Toulouse, France
  \and
Dipartimento di Fisica e Astronomia "Galileo Galilei", Università di Padova, Vicolo dell'Osservatorio 3, Padova, Italy, IT-35122
  }
   
 
  \abstract
   {}
   {The source X 1822-371 is an eclipsing compact binary system with a period close to 5.57 hr and an orbital period derivative $\dot{P}_{\rm orb}$ of 1.51(7)$\times 10^{-10}$ s s$^{-1}$. The very large value of $\dot{P}_{\rm orb}$ is compatible with a super-Eddington mass transfer rate from the companion star, as suggested by X-ray and optical data. The \textit{XMM-Newton} observation taken in 2017 allows us to update the orbital ephemeris and verify whether the orbital period derivative has been stable over the last 40 yr.}
   {We added to the X-ray eclipse arrival times from 1977 to 2008 two new values obtained from the \textit{RXTE} and {\it XMM-Newton} observations performed in 2011 and 2017, respectively. We estimated the number of orbital cycles and the delays of our eclipse arrival times spanning 40 yr using as reference time the eclipse arrival time obtained from the \textit{Rossi-XTE} observation taken in 1996.}
   {Fitting the delays with a quadratic model, we found an orbital period $P_{\rm orb}=5.57062957(20)$ hr and a $\dot{P}_{\rm orb}$ value of $1.475(54) \times 10^{-10}$ s s$^{-1}$. The addition of a cubic term to the model does not significantly improve the quality of the fit. We also determined a spin-period  value of $P_{\rm spin}=0.5915669(4)$ s and its first derivative $\dot{P}_{\rm spin}= -2.595(11) \times 10^{-12}$ s s$^{-1}$.}
  {The obtained results confirm the scenario of a super-Eddington mass transfer rate; we also exclude a gravitational coupling between the orbit and the change in the oblateness of the companion star triggered by the nuclear luminosity of the companion star.}

  \authorrunning{S. M. Mazzola et al.}

  \titlerunning{Updated orbital ephemeris of X 1822-371 spanning 40 years}
  
  \keywords{stars: neutron -- stars: individual: X 1822-371  ---
  X-rays: binaries  --- eclipses, ephemerides}
  

   \maketitle
%

\section{Introduction}
The low-mass X-ray binary system (LMXB) X 1822-371 is a persistent eclipsing source with an orbital period of 5.57 hr, hosting an accreting X-ray pulsar with a spin frequency close to 1.69 Hz \citep{jonker_01}, that is increasing with a derivative of $\dot{\nu}=(7.57 \pm 0.06) \times 10^{-12}$ Hz s$^{-1}$ \citep{baknielsen_17, iaria_15}. The mass function of the system is $(2.03 \pm 0.03) \times 10^{-2} \ \rm{M}_{\odot}$ \citep{jonker_01}, with a lower limit on the companion star mass of $0.33 \pm 0.05 \ \rm{M}_{\odot}$ \citep{jonker_03}. X 1822-371 belongs to the class of accretion disc corona (ADC) sources \citep{white_82}, with an inclination angle between \ang{81} and \ang{84} \citep{heinz_01}. 
The distance to this source was estimated to be between 2-2.5 kpc by \cite{mason_82} using infrared and optical observations. The 0.1-100 keV unabsorbed luminosity is $1.2 \times 10^{36}$ erg s$^{-1}$, adopting a distance of 2.5 kpc \citep{iaria_01}.
The most recent orbital ephemeris of the source X 1822-371 was reported by \cite{Chou_16}, who suggested that the orbital period derivative is $\dot{P}_{\rm orb}=(1.464 \pm 0.041) \times 10^{-10}$ s s$^{-1}$ adopting quadratic ephemeris, or $\dot{P}_{\rm orb}=(1.94 \pm 0.27) \times 10^{-10}$ s s$^{-1}$ adopting cubic ephemeris. The value of $\dot{P}_{\rm orb}$ is three orders of magnitude larger than what is expected from conservative mass transfer driven by magnetic breaking and gravitational radiation and can be explained only by assuming a mass transfer rate larger than three times the Eddington limit for a neutron star \citep{burderi_10, bay_10}. \cite{baknielsen_17} suggested that X 1822-371 is a relatively young binary in whichthe donor is transferring mass on a  thermal time-scale. The authors suggested that the super-Eddington mass transfer rate generates an outflow of matter from the magnetospheric radius.

A suggestion to explain the evolutionary stage of X1822-371 comes also from recent numerical studies of the secular evolution of LMXBs including X-ray irradiation of the donor \citep{tailo_18}. These models show that when the donor has a mass 0.4$\lesssim$ M/M$_\odot \lesssim$0.6, like in this system, the evolution is subdivided into cycles of short mass transfer phases, during which the donor expands on the thermal timescale of its convective envelope and the orbital period increases significantly, followed by long phases of detachments during which thermal relaxation takes place and the donor recovers full thermal equilibrium. The next stage of mass transfer occurs when the orbital period has decreased again so that the stellar radius fills again the Roche lobe, and a new orbital expansion follows. The maximum  $\dot{P}_{\rm orb}$ in the published models is $\sim 6 \times 10^{-11}$ s s$^{-1}$ \citep[see e.g.][]{tailo_18}, but the specific evolution of X1822-371 may be obtained by reasonable variations of the input parameters.

In this work, we used the eclipse arrival times reported by \cite{iaria_11}, with the addition of two new eclipse arrival times obtained analysing the \textit{Rossi-XTE} observation performed in 2011 and the \textit{XMM-Newton} observation performed in 2017; our eclipse arrival times span 41 yr. 
We investigated the statistical significance for the presence of a second derivative of the orbital period and the possibility that the quadratic term mimics a wide sinusoidal modulation. In the latter case, we excluded that the sinusoidal modulation could be explained as due to a gravitational coupling of the orbit with changes in the oblateness of the magnetically active companion star, the so-called Applegate mechanism \citep{apple_92}.
 \begin{figure}[htbp!]
 \centering
      \includegraphics[width=7.cm]{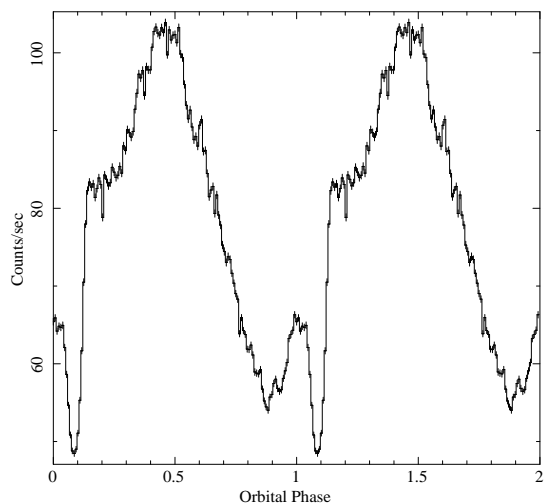}
      \caption{{\it XMM-Newton/Epn} folded orbital light curve obtained adopting a period of 0.2321107 days. The period is divided in 128 bins.}
        \label{fig:epn_lc_corr}
    \end{figure}
    
\section{Observations and Data Analysis}
   The \textit{XMM-Newton} Observatory \citep{jansen_01} observed the source X 1822-371 in 2017 March 3rd between 01:10:54 UTC and 19:12:27 UTC (ObsId. 0784820101) for a total observing time of 69 ks.
   We analysed the data collected by pn-type CCD detector of the European Photon Imaging Camera \citep[Epn,][]{struder_01}, operating in Timing Mode, with the aim to estimate the eclipse arrival time.
   We reprocessed the data using the Science Analysis Software (SAS) v16.1.0, verified the absence of background flaring during the observation and applied the barycentric correction to the event times.
 
    We extracted the Epn 0.3-10 keV light curve considering only PATTERN$\leq$4 and FLAG=0 events from a region which included the brightest columns of the detector (RAWX between 30 and 45) while for the background we extracted the events from a region far away from the source (RAWX between 5 and 10). 
    We observe three partial eclipses at 14 ks, 34 ks and 54 ks from the start time in the Epn background-subtracted light curve. The observation covers almost 3.4 orbital periods of the system.
 \begin{table}[!htbp]
        \centering
        \caption{Journal of available eclipse arrival times for the source X 1822-371}
        \scriptsize
        \begin{tabular}{ccccc}
        \hline
        \hline
        Eclipse time & Delays & Cycle & Ref. & Satellite \\  
        (MJD, TDB) & (s) &  &  \\  
        \hline \\
        43413.0272(46) & 1416(397) & -29900 & 1 & HEAO-1 Scan\\
        43591.0521(46) & 1145(397) & -29133 & 1 & HEAO-1 Scan\\
        43776.0459(12) & 1359(104) & -28336 & 1 & HEAO-1 Point\\
        43777.9065(46) & 1680(397) & -28328 & 1 & HEAO-1 Scan\\
        43968.9247(69) & 991(596) & -27505 &  2 & Einstein\\
        44133.0277(30) & 1124(259) & -26798 & 1 & Einstein\\
        45579.9932(5) & 642(43) & -20564 & 1 & EXOSAT\\
        45614.80940(38) & 622(33) & -20414 & 1 & EXOSAT\\
        45962.50914(33) & 588(29) & -18916 & 1 & EXOSAT\\
        45962.74046(30) & 520(26) & -18913 & 1 & EXOSAT\\
        45962.97254(54) & 517(29) & -18912 & 1 & EXOSAT\\
        46191.13643(31) & 533(27) & -17929 & 1 & EXOSAT\\
        46191.36768(33) & 459(29) & -17928 & 1 & EXOSAT\\
        46191.60008(29) & 484(25) & -17927 & 1 & EXOSAT\\
        47759.72810(30) & 195(26) & -11171 & 1 & Ginga\\
        48692.34396(70) & 83(60) & -7153 & 1 & ROSAT\\
        49267.50984(40) & -58(35) & -4677 & 3 & ASCA\\
        50352.85425(35) & -54(30) & -1 & 3 & ASCA\\
        50353.08728(23) & 26(20) & 0 & 3 & RXTE\\
        50701.0187(12) & 46(104) & 1499 & 3 & BeppoSAX\\
        50992.0858(23) & 101(199) & 2753 & 4 & RXTE\\
        51779.6317(19) & -61(164) & 6146 & 4 & Chandra\\
        51975.06934(56) & 59(48) & 6988 & 4 & XMM-Newton\\
        51975.06935(31) & 59(27) & 6988 & 4 & RXTE\\
        52432.09458(30) & 188(26) & 8957 & 4 & RXTE\\
        52488.03300(38) & 189(33) & 9198 & 4 & RXTE\\
        52519.13569(85) & 190(73) & 9332 & 4 & RXTE\\
        52882.15470(37) & 158(32) & 10896 & 4 & RXTE\\
        54010.6730(9) & 294(78) & 15758 & 5 & Suzaku\\
        54607.19592(56) & 408(48) & 18328 & 4 & Chandra\\
        55887.05307(38) & 838(33)& 23842 & 6 & RXTE \\
        57818.44392(96) & 1452(82) & 32163 & 6 & XMM-Newton\\
\hline
\hline
\end{tabular}
    \footnotesize{
    
    \textbf{References}:1) \cite{hellier_94}, 2) \cite{hellier_89}, 3) \cite{parmar_00}, 4) \cite{burderi_10}, \\ 5) \cite{iaria_11}, 6) this work}
        \label{tab:time}
    \end{table}
    We folded the background-subtracted light curve, adopting a reference epoch $T_{\rm fold}=57818.4237$ MJD (corresponding to a time close to the mid-time of the observation) and a reference period of $P_{\rm fold}=0.2321107$ days. The folded light curve is shown in \autoref{fig:epn_lc_corr}.
    
    We further added the X-ray eclipse time obtained by analysing the \textit{RXTE} observations taken from 2011 November 15 to 30 (ObsId. P96344). The same observation was analysed by \cite{Chou_16} using standard 2 data in the energy range 2-9 keV and inferring four eclipse arrival times. In order to make the analysis self-consistent, we re-analysed these data using the X-ray light curves obtained from the Standard 1 data products (Std1) of archival \textit{RXTE} data, that is the 2-40 keV background-subtracted light curves collected by the PCA with a time resolution of 0.125 s. We applied the barycentric correction to the events using the ftool {\tt faxbary} and folded the light curve using as epoch $T_{\rm fold}=55887$ MJD and as period $P_{\rm fold}=0.2321104$ days. 
    To estimate the orbital phase at which the eclipse occurs we adopted the procedure reported by \cite{burderi_10} finding that the eclipse arrival times are $T_{\rm ecl}=57818.44392(96)$ MJD/TDB and $T_{\rm ecl}=55887.05307(38)$ MJD/TDB for the \textit{XMM-Newton/Epn} and \textit{RXTE/PCA} observations, respectively. The associated errors are at 68\% confidence level.
    \begin{table}[!htbp]
        \centering
        \caption{Best-fit parameters of the modeling of eclipse time delays with different models including quadratic, cubic, sinusoidal, and quadratic plus sinusoidal ephemeris.}
        \scriptsize
        \begin{tabular}{lcc}
        \hline
        \hline
        Parameter & Quadratic & Cubic  \\
                  & Model     & Model   \\
    \hline
    $a$ (s)  &  $5 \pm 15$ &  $-3 \pm 14$   \\
    $b$ ($10^{-4}$ s) &  $-5 \pm 7$&  $19 \pm 16$ \\
    $c$ ($10^{-6}$ s) &$1.48 \pm 0.05$& $1.52 \pm 0.06$ \\
    $d$ ($10^{-12}$ s)  &  --  &  $-6 \pm 4$\\
    $T_{\rm {0,orb}}$ (MJD/TDB) & $50353.08733(16)$ & $50353.08725(16)$ \\
    $P_{\rm 0,orb}$ (days) &  $0.2321095653(85)$ & $0.232109593(18)$  \\
    $\dot{P}_{\rm orb}$ ($10^{-10} $ s s$^{-1}$) & $1.475(54)$
    & $1.514(55)$ \\ 
    $\Ddot{P}_{\rm orb}$ ($10^{-19} $ s s$^{-2}$) & --& $-0.91(55)$ \\
    $\chi^2$/d.o.f. & 42.3/29  & 37.4/28 \\
    \hline
          Parameter &  LS & LQS\\
                   & Model& Model \\
    \hline
    $a$ (s)  & $8084$ (fixed) & $4 \pm 13$ \\
    $b$ ($10^{-4}$ s) & $-873.9$ (fixed) & $-4 \pm 7$\\
    $c$ ($10^{-6}$ s)  & -- & $1.47 \pm 0.05$ \\
    $A$ (s) &  $9290 \pm 30$ & $34\pm 12$\\
    $a_{\rm bin}/l$ & --& $1.1\pm 0.3$\\
    N$_{\rm MOD}$ ($ \times 10^4$) & $32.1 \pm 0.5$ &  $0.5 \pm 0.2$   \\
    N$_0$  ($ \times 10^4$) & $5.41\pm 0.06$ & $-0.28 \pm 0.03$  \\
    P$_{\rm MOD}$ (yr) & $204 \pm 3$ & $3.4 \pm 1.2$\\
    $T_{\rm {0,orb}}$ (MJD/TDB) & $50353.18085$ (fixed) & $50353.08733(15)$\\
    $P_{\rm 0,orb}$ (days) & $0.232108660$ (fixed) & $0.2321095661(81) $\\
    $\dot{P}_{\rm orb}$ ($10^{-10} $ s s$^{-1}$)  & --& $1.468(53)$\\ 
    $\chi^2$/d.o.f.  & 37.8/29 & 31.2/26\\ 
 \hline   
\hline
\end{tabular}
\tablefoot{The errors are at 68\% confidence level} 
        \label{tab:parameter}
    \end{table}

  \begin{figure*}[!htbp]
   \centering
    \includegraphics[scale=1.1]{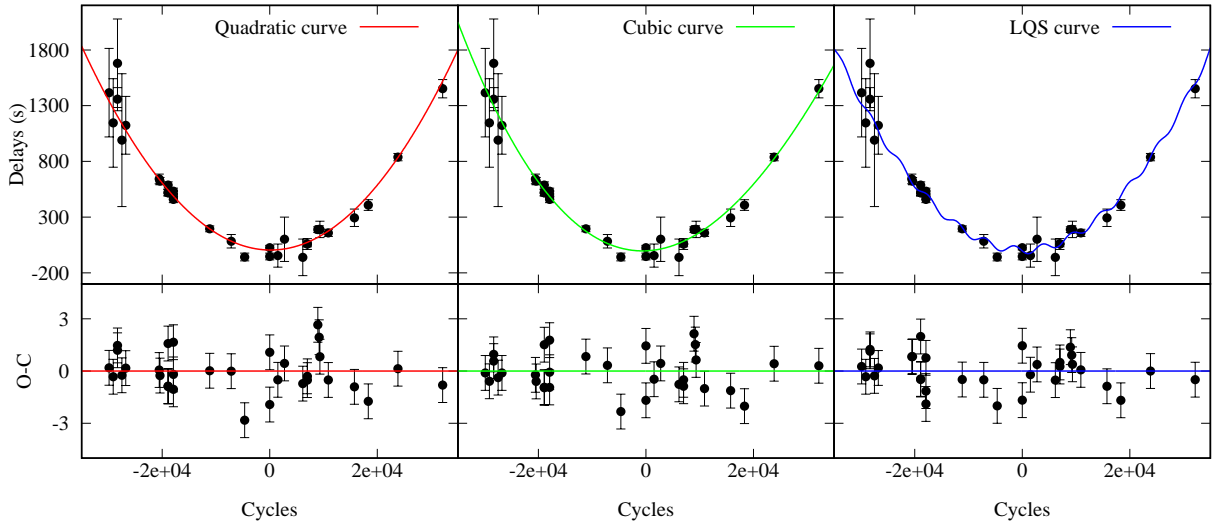}
    \caption{From the left to the right: delays vs. cycles for quadratic (red), cubic (green) and LQS (blue) model. Residuals are in units of $\sigma$ obtained adopting the quadratic, cubic and LQS model, respectively.}
   \label{fig:delay_tidal}
\end{figure*}
    To update the orbital ephemeris we included the two eclipse arrival times shown above to those reported by \cite{iaria_11}; the 32 eclipse arrival times, the corresponding number of orbital cycles and the delays are summarised in \autoref{tab:time}. 
    The number of orbital cycles $N$ and the delays associated to the eclipse arrival times were obtained adopting a reference orbital period of $P_{\rm 0}=0.232109571$ days and a reference eclipse time $T_0=50353.08728$ MJD, estimated for the {\it RXTE} observation of the source performed in 1996 \citep{parmar_00}.
\begin{figure*}
    \centering
    {\includegraphics[scale=0.55]{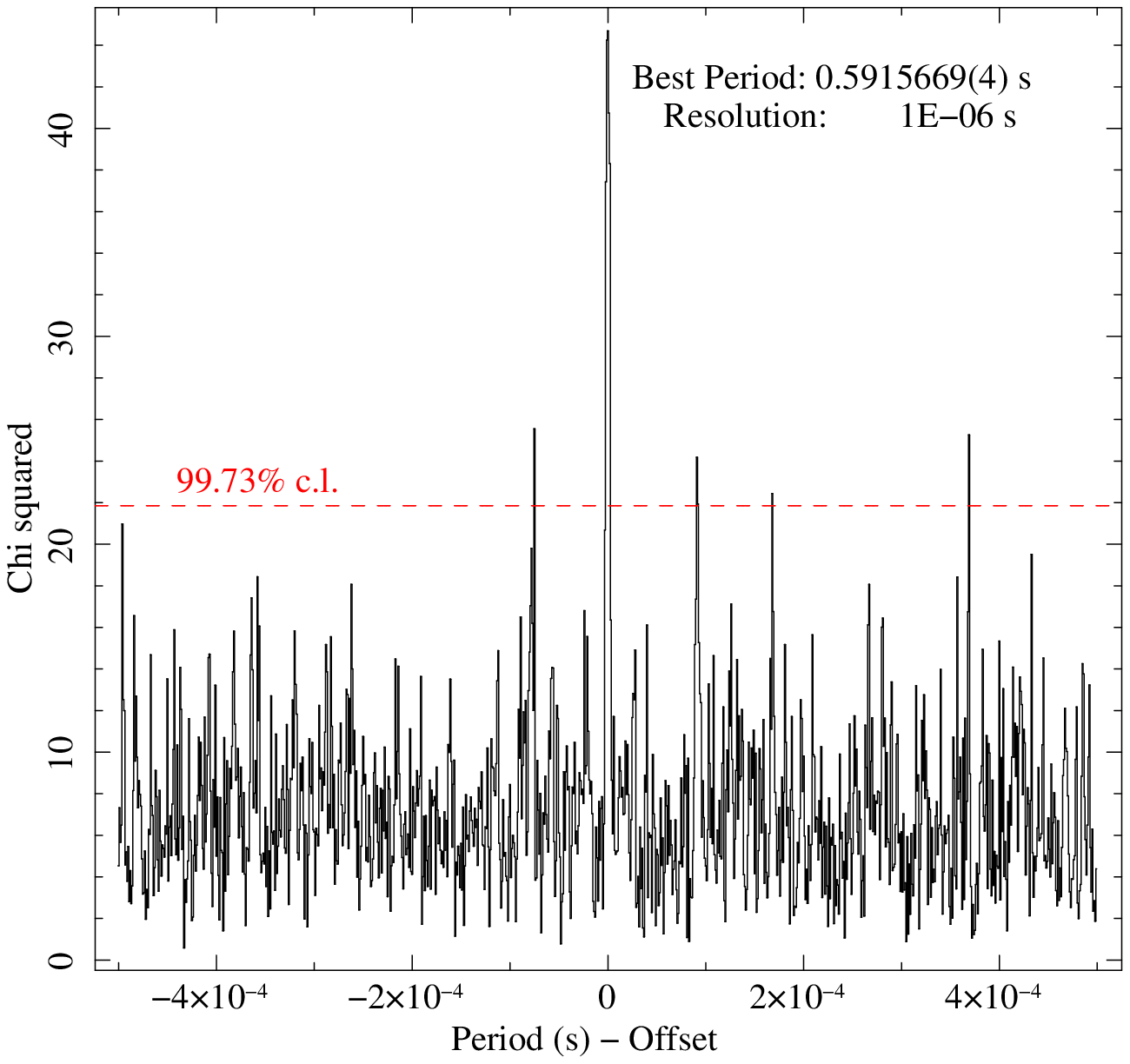}} \hspace{2cm}
    {\includegraphics[scale=0.55]{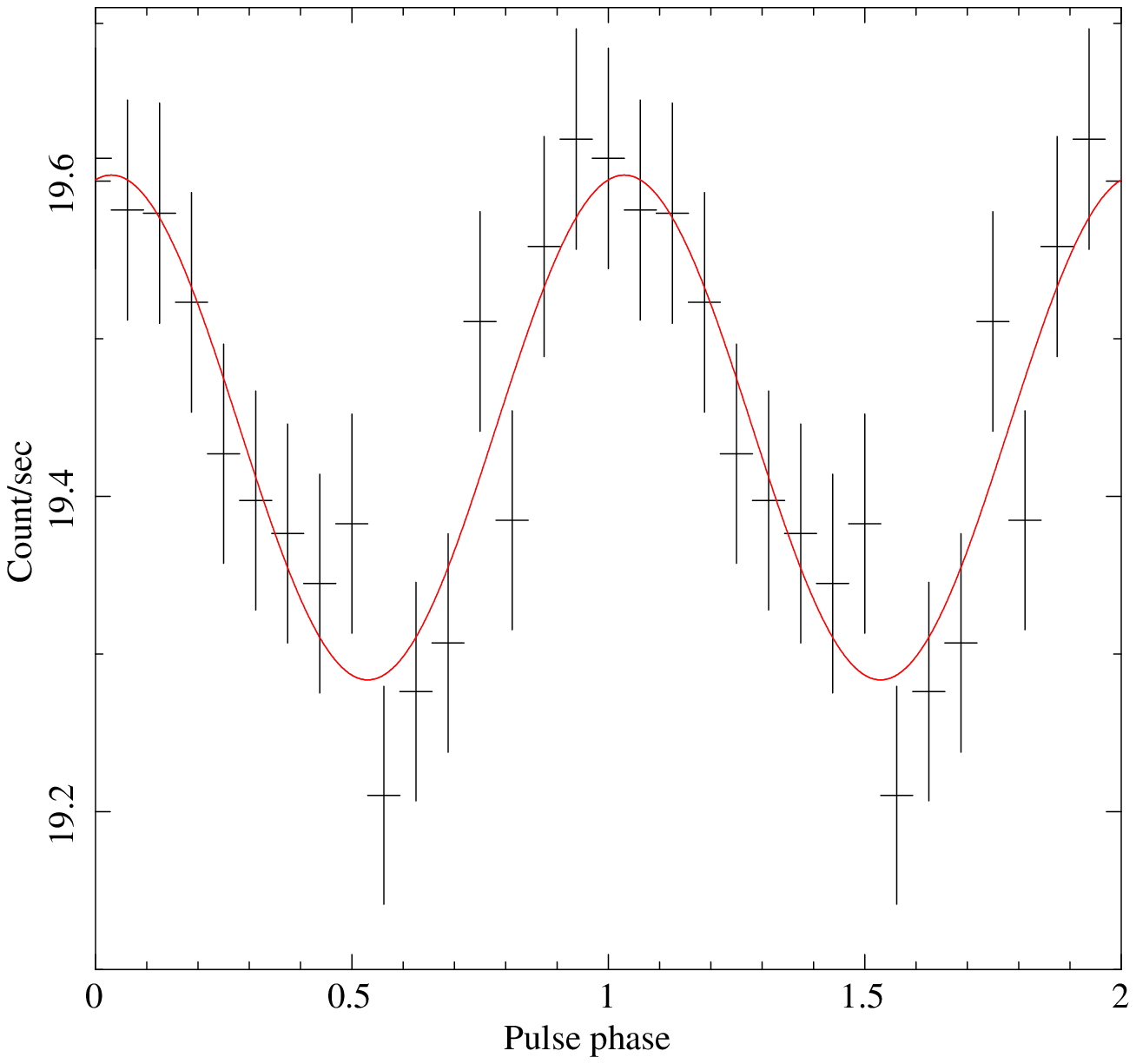}}
    \caption{Left panel: Folding search for periodicity in the 5-12 keV \textit{XMM-Newton/Epn} light curve. The horizontal dashed line indicate the $\chi^2$ value of 21.85 at which we have the 99.73\% confidence level for a single trial, corresponding to a significance of 3 $\sigma$.
    Right panel: \textit{XMM-Newton/Epn} folded light curves obtained adopting the best period and using 16 phase bins per period.
    }
    \label{fig:pulse}
\end{figure*}
  
     We fitted the delays as function of cycles adopting the quadratic model $y=a+bN+cN^2$ and obtaining a $\chi^2(d.o.f.)$ of 42.3(29). The uncertainties associated with the best-fit parameters $a$, $b$ and $c$ have been scaled by a factor $(\chi^2_{\rm red})^{1/2}$ to take into account a $\chi^2_{\rm red}$ of the best-fit model larger than 1. The best-fit values of the parameters are shown in the second column of \autoref{tab:parameter} (upper part); the best-fit quadratic curve (red color) and the corresponding residuals in units of $\sigma$ are shown in the left panel of \autoref{fig:delay_tidal}.

    The updated orbital ephemerides are:
\begin{align}
\centering
        T_{\rm ecl}= 50353.08733(16) \; {\rm MJD/TDB} + 0&.2321095653(85) N \\ \notag
       &+ 1.711(63)\times 10^{-11} N^2,
\end{align}
where the first and the second term represent the new values of the reference epoch $T_{\rm 0,orb}$ and orbital period $P_{\rm 0,orb}$, respectively. The third term, equal to $\left( P_{\rm 0} \dot{P}_{\rm orb}\right)/2$, allows us to estimate an orbital period derivative of $\dot{P}_{\rm orb}= 1.475(54) \times 10^{-10}$ s s$^{-1}$.
Furthermore, we added a cubic term $d=\left(P^2_{\rm 0} \ddot{P}_{\rm orb}\right)/6$ to the quadratic model in order to test the presence of a second derivative of the orbital period. The best-fit curve (green color) and the corresponding residuals are shown in \autoref{fig:delay_tidal} (central panel); the best-fit parameters are shown in the third column of \autoref{tab:parameter} (upper part). We obtained a $\chi^2(d.o.f.)$ of 37.4(28), the F-test probability of chance improvement is 0.065 indicating that the cubic model improves the fitting at a confidence level less than 2$\sigma$, meaning that the cubic term is not significantly required. 

We investigated also whether the quadratic term could mimic a sinusoidal modulation in the delays: we substituted the quadratic term with a sinusoidal one, using the model 
$y=a+bN+A\sin[2\pi(N-N_0)/N_{\rm MOD}]$
composed of a linear plus sinusoidal term (LS model, hereafter).
Keeping fixed the best-fit values of $a$ and $b$ to 8084.33 s and -0.0873931 s to lead the fit to the convergence, we obtained a $\chi^2(d.o.f.)$ of 37.8(29) with a $\Delta \chi^2$ of 4.9 with respect to the quadratic model, a modulation period $P_{\rm MOD}=N_{\rm MOD} \; P_{\rm 0}=204 \pm 3$ yr and a semi-amplitude of the modulation $A=9289(26)$ s. The best-fit values are shown in the second column of the lower part of \autoref{tab:parameter}. We also verified whether a gravitational qu-adrupole coupling produced by tidal dissipation \citep{apple_94} could be detectable in our data; to this aim we added a quadratic term to the LS model. In the new model (hereafter LQS model) we imposed that $N_{\rm MOD} = 0.572 \ c^{-1/2} A^{2/3} a_{\rm bin}/l$ (this relation will be discussed in Section 3). The best-fit parameters are shown in the third column of the lower part in \autoref{tab:parameter}; the best-fit model (blue color) and the corresponding residuals are shown in the right panel of \autoref{fig:delay_tidal}. We obtained a $\chi^2(d.o.f.)$ of 31.2(26), the F-test probability of chance improvement is 0.045 with respect to the quadratic model indicating that LQS model improves the fit at a confidence of about 2$\sigma$.

Finally, we looked for the presence of the NS spin frequency in the \textit{XMM-Newton/Epn} data, analysing the 5-12 keV events after applying the barycentric correction using the source coordinates. We corrected the data for the binary orbital motion using $a \sin i=1.006(5)$ lt-s \citep{jonker_01} and the value of the orbital period obtained from the quadratic ephemeris shown above. In order to search for the pulsation period, we applied the procedure described by \cite{iaria_15}: we used the ftool {\tt efsearch} of the XRONOS package (v 5.22), adopting as reference time the start time of the observation and a resolution of the period search of $ 10^{-6}$ s. We explored around a period $P_{\rm spin}$ of 0.591567 s, estimated using Eq. 6 in \cite{Chou_16}, and subsequently, we fitted the peak of the corresponding $\chi^2$ curve with a Gaussian function. We assumed that the centroid of the Gaussian was the best estimation of the spin period and we associated to this value the 68\% c. l. error obtained from the best fit.
We found the spin period is 0.5915669(4) s, the $\chi_{\rm peak}^2$ associated to the best period is 44.66 (see the left panel in \autoref{fig:pulse}) and the probability of obtaining a $\chi^2$ value greater than or equal to $\chi_{\rm peak}^2$ by chance, having seven degrees of freedom, is $1.58 \times 10^{-7}$ for a single trial. Considering the 1000 trials in our research, we expect almost $1.58 \times 10^{-4}$ periods with a $\chi^2$ value greater than or equal to $\chi_{\rm peak}^2$. This implies a detection significance at the 99.984\% confidence level.

Furthermore, we folded the 5-12 keV \textit{XMM-Newton/Epn} light curve adopting the obtained $P_{\rm spin}= 0.5915669(4)$ s and the start time of the observation as reference epoch; we utilised 16 phase bins per period. We fitted the folded light curve with a constant plus a sinusoidal function with period kept fixed to one and we obtained a $\chi^2(d.o.f)$ of 9.176(12), a constant value of $19.44(2)$ c/s and a sinusoidal amplitude $A= 0.16(2)$ c/s. We show the folded light curve and the best-fit curve in the right panel of \autoref{fig:pulse}.
We found the fractional amplitude of the pulsation to be $0.83 \pm 0.11\%$ for the estimated background count rate of $0.15(1)$ c/s. This value is compatible with that reported by \cite{jonker_01} in the 5-12 keV energy band.

In the end, using the spin period values reported by \cite{iaria_15} in Tab. 2, the last value reported by \cite{Chou_16} in Tab. 4 and the spin-period obtained above, we estimated a spin period derivative of $-2.595(11) \times 10^{-12}$ s s$^{-1}$ with
\begin{align}
\centering
P_{\rm spin}(t) = 0.592&758(3) \ {\rm s} \\ \notag
&- 2.595(11) \times 10^{-12} \ (t -52500 \ {\rm MJD}) \times  86400.
\end{align}

\section{Discussion}
We updated the orbital ephemeris of the source X 1822-371 by adding two eclipse arrival times obtained from the \textit{RXTE/PCA} observations performed in 2011 and from the \textit{XMM-Newton/Epn} observation performed in 2017. Our baseline covers almost 40 yr, from 1976 to 2017. We moved the reference epoch of the ephemeris to a more recent time, corresponding to 1996 Sep. 27, that is close to the middle of the baseline. 
 We inferred a $\dot{P}_{\rm orb}$ of $1.475(54) \times 10^{-10}$ s s$^{-1}$, compatible with the values present in literature. We explored the possibility that a cubic model could improve the fit of the delays as suggested by \cite{Chou_16}; the addition of a cubic term to the quadratic model does not improve significantly the fit yet. 

Several authors \citep{burderi_10,bay_10,baknielsen_17} explained the large value of $\dot{P}_{\rm orb}$ as due to a super-Eddington non-conservative mass transfer rate. This ‘quadratic model’ seems the simplest explanation. We alternatively investigated the possibility that a large sinusoidal modulation could mimic the quadratic trend of the delays. A sinusoidal modulation of the delays could be associated to the gravitational quadrupole coupling (GQC) between the orbit and the changes of the quadrupole moment of the magnetically active companion star \citep{apple_92}. The magnetic activity of the secondary generates a torque to the subsurface magnetic field of the companion star (CS); the torque induces a cyclic exchange of angular momentum between the inner and outer regions of the CS changing its gravitational quadrupole moment and, consequently, the orbital period of the binary system. 
We assumed that the necessary luminosity $L_{\rm GQC}$ to activate this mechanism comes from the nuclear luminosity $L_{\rm nuke}$ produced by the CS itself \citep{apple_92}. We assumed the mass function $f=(2.03 \pm 0.03) \times 10^{-2} \ M_{\odot}$ \citep{jonker_01} and the inclination angle $82.5 \pm 1.5$ deg \citep{heinz_01}, then we estimated the mass ratio $q=M_2/M_1=0.27 \pm 0.02$ adopting a CS mass $M_2$ of $0.46 \pm 0.02 \ M_{\odot}$ and a NS mass $M_1$ of $1.69 \pm 0.13 \ M_{\odot}$ \citep{iaria_15}. Under the reasonable hypothesis that the CS fills its Roche lobe, using eq. 15 in \cite{Sanna_17}
\begin{equation}
L_{\rm GQC} = 3.35 \times 10^{32}m_1 q^{1/3} (1+q)^{4/3} P^{-2/3}_{\rm orb, 5h} \frac{A^2}{P^3_{MOD,yr}} {\; \rm erg \; s^{-1}},
\end{equation}
where $m_1$ is the NS mass in units of M$_{\odot}$, $A$ the semiamplitude of the sinusoidal modulation in seconds, $P_{\rm MOD,yr}$ the modulation period in yr and $P_{\rm orb, 5h} $ the orbital period in units of five hr, we inferred that $L_{\rm GQC}=(2.14 \pm 0.22) \times 10^{33}$ erg s$^{-1}$ adopting the best-fit values of $A$ and $P_{\rm MOD,yr}$ obtained from the LS model.
The nuclear luminosity of a star with mass $0.43 \ M_{\odot} < M < 2\ M_{\odot}$ is given by $L_{\rm nuke}/L_{\odot}= m^{4}$, where $m$ is the stellar mass in units of solar masses \citep{sal_05}. 
Substituting to the latter expression the value of $m_2$, we find that $L_{\rm nuke}= (1.71 \pm 0.14) \times 10^{32}$ erg s$^{-1}$, implying that the nuclear luminosity is a factor of 13 lower than the luminosity needed to activate the GQC process. Hence, a large sinusoidal modulation in the delays cannot be explained as results of an Applegate mechanism powered by the nuclear energy of the companion.
It is more reasonable that the delays follow a quadratic trend caused by a the high value of the orbital period derivative.

It could be possible on the other hand that the energy transferred to the CS to trigger the GQC process occurs via tidal dissipation \citep[][for a discussion]{apple_94,Sanna_17}. In this scenario, the magnetic field of the CS, interacting with the mass ejected from the system because of the irradiation from the accreting neutron star, could produce a torque able to slow down the rotation of the CS. The torque, then, holds the CS out of synchronous rotation generating a tidal dissipation that could furnish the necessary energy to activate the GQC process. In this case, the CS should lose mass and therefore we should observe an orbital period derivative. Combining the eqs. 17 and 18 of \cite{Sanna_17} we find that the mass transfer rate to trigger the GQC process via tidal dissipation is:
\begin{equation}
\label{eq:mdot_tidal}
\dot{m}_T = 1.415 \times 10^{-8} \left( \frac{a_{\rm bin}}{l} \right)^2 m_1^{11/9} \frac{q^{7/9}}{(1+q)^{1/9}}\frac{A^{4/3}}{P_{MOD,yr}^2} \ {\rm M_{\odot} \ yr^{-1}},
\end{equation}
where $a_{\rm bin}/l$ represents the ratio between the binary separation and the lever arm of the mass transferred by the CS measured with respect to the center of mass of the binary system. On the other hand, the mass transfer rate from the CS is linked to the $P_{\rm orb}$ and $\dot{P}_{\rm orb}$ values as reported in eq. 4 of \cite{burderi_10}, that is
\begin{equation}
\label{eq:mdot_transf}
\dot{m} = 0.39 \ (1-3n)^{-1} m_2 \ c \ P^{-1}_{\rm orb, 5h} \ {\rm M_{\odot} \ yr^{-1}},
\end{equation}
where $n$ is the mass-radius index of the CS and $c$ is the constant of the quadratic term in the model adopted to fit the delays. We assumed $n=-1/3$ as reported by \cite{burderi_10}. 
Combining the Equations \ref{eq:mdot_tidal} and \ref{eq:mdot_transf} we obtained $N_{\rm MOD} = 0.572 \ c^{-1/2} A^{2/3} a_{\rm bin}/l$, which we used to constrain the best-fit model (see Section 2).
The best-fit values obtained from this LQS model suggest that the GQC process is possible via tidal interaction if the mass transfer rate is $(9.4 \pm 0.3) \times 10^{-8}$ M$_{\odot}$ y$^{-1}$, that is a factor of 6 larger than the Eddington mass accretion rate. 

\section{Conclusions}
Our results confirm the scenario of a super-Eddington mass transfer rate for X 1822-371, where most of the transferred mass is expelled from the system by the X-ray radiation pressure and only a fraction of it accretes onto the neutron star \citep[see e.g.][]{Iaria_13}.
Apart from this simplest ‘quadratic model’, we note that the GQC mechanism via tidal interaction also predicts a parabolic trend of the delays over which a small modulation with a period of $3.4\pm1.2$ yr and an amplitude of $34\pm12$ s is superimposed. In other words, both the quadratic model and the GQC mechanism via tidal interaction require a large outflow of mass,  several  times  the Eddington limit, from the system.

\begin{acknowledgements}
 The authors thank Dr. Francesca D’Antona for the helpful and kind discussion.
The authors acknowledge financial contribution from the agreement ASI-INAF n. 2017-14-H.0, and from the HERMES Project, financed by the Italian Space Agency (ASI) Agreement n. 2016/13 U.O.
Part of this work has been funded by the research grant “iPeska” (PI: Andrea Possenti) funded under the INAF national call Prin-SKA/CTA approved with the Presidential Decree 70/2016.
The authors would like to thank the anonymous Referee for
his/her helpful comments.
\end{acknowledgements}
\bibliographystyle{aa}
\bibliography{biblio}
\end{document}